# ПРОЕКТУВАННЯ КОМБІНОВАНОГО НАВЧАННЯ СИСТЕМНОГО ПРОГРАМУВАННЯ БАКАЛАВРІВ ПРОГРАМНОЇ ІНЖЕНЕРІЇ

А. М. Стрюк

м. Кривий Ріг, ДВНЗ «Криворізький національний університет»
andrey.n.stryuk@gmail.com

Сутність педагогічного проектування визначається в діяльності викладача ВНЗ як спрямованість на відбір змісту навчальної діяльності з метою формування у студентів предметних спеціальних знань і вмінь, на розробку ефективних технологій, діагностування результатів проектної педагогічної діяльності на основі об'єктивних критеріїв готовності майбутніх фахівців до професійної діяльності [6].

В. І. Гриценко визначає педагогічне проектування як «розробку засобів подання навчального матеріалу, вибір телекомунікаційних та інформаційних технологій і формування навчальних стратегій ... з урахуванням цілей навчання і стилю вивчення потенціальної аудиторії» [5, 156]. Нашою задачею є проектування комбінованого навчання системного програмування бакалаврів програмної інженерії. Під комбінованим навчаннями ми розуміємо спосіб реалізації змісту навчання, що інтегрує аудиторну та позааудиторну навчальну діяльність за умови педагогічно виваженого поєднання технологій традиційного, електронного, дистанційного та мобільного навчання з метою ефективного досягнення навчальних цілей.

Реалізація комбінованого навчання потребує розробки комплексу навчально-методичних матеріалів, що включають в себе робочу програму курсу, навчальні та інструктивні матеріали, перелік практичних, лабораторних робіт і методичних порад щодо їх виконання, план проведення та тематику дискусій, семінарів та інших форм навчальної діяльності, передбачену курсом.

В сучасній педагогічній практиці існує багато підходів до проектування та розробки навчальних матеріалів та систем навчання. Широке розповсюдження отримала модель розробки систем навчання ADDIE (Analysis, Design, Development, Implementation, and Evaluation) [4], згідно якої розробка та використання навчальних матеріалів має складатися з п'яти етапів: *аналізу* – визначення цілей навчання і завдання, які має виконати той, хто навчається, щоб продемонструвати та застосувати отриманні знання та навички; *проектування* – конкретизації цілей і завдань окремих розділів курсу, планування занять, визначення структури навчальних матеріалів та засоби, що будуть використовуватись під час навчання; *розробка* – створення необхідних навчальних матеріалів, інтеграція засобів ІКТ, друк або розміщуються в системі управління навчанням; *реалізація* – безпосередньо здійснення навчального процесу за розробленою програмою з використанням створених матеріалів; *оцінка* – аналіз результатів навчання, за результатами якого вносяться зміни до відповідних навчальних матеріалів та планів.

Р. Райзер та Дж. Демпсі уточнюють модель ADDIE і зазначають, що «область педагогічного проектування ... включає в себе аналіз навчальних та практичних задач, проектування, розробку, впровадження, оцінювання та управління навчальними та супровідними процесами та ресурсами, що залучені для поліпшення навчального процесу» [3, 5]. Дж. Кемп [2] підкреслює, що процес навчання є неперервним циклом, що потребує постійного планування, проектування, розробки і оцінки і пропонує модель, що нараховує дев'ять етапів: аналіз цілей, аналіз аудиторії, аналіз завдань, деталізація цілей, структурування матеріалу, розробка плану, розкладу, реалізація навчання та оцінювання. Подібні етапи виділяє і О. Е. Коваленко [7] в функціональній схемі управління процесом навчання (рис. 1), яка більш системно відображає зв'язки між різними етапами процесу навчання, а також підкреслює важливу роль технологічної складової, що передбачає обґрунтований вибір засобів навчання, а також можливість застосування додаткових засобів, якщо процес навчання потребує корегування.

Перш за все сформулюємо загальну ціль навчання: сформувати у студентів бакалаврату з програмної інженерії компетенції зі створення системного програмного забезпечення та використання системних викликів та сервісів операційних систем при розробці прикладного програмного забезпечення. Аналіз Галузевого стандарту вищої освіти, SWEBOK [1] та типових посадових інструкцій інженера програміста і системного програміста дозволив визначити дії, пов'язані з системним програмуванням, які бакалавр програмної інженерії має виконувати під час практичної діяльності:

– проектування програмного забезпечення;

– аналіз, проектування та прототипування людино-машинного інтерфейсу;

– розробка алгоритмів та структур даних для програмних продуктів;

– на основі аналізу математичних моделей і алгоритмів розробка програми, її тестування та налагодження;

– визначення обсягу, структури, схеми введення, опрацювання, збереження і виведення даних, що підлягають обробці засобами обчислювальної техніки;

– конструювання інструментального програмного забезпечення для розробки системного та прикладного програмного забезпечення (компіляторів, текстових процесорів, оболонок операційних систем);

– конструювання операційних систем та їх оточення;

– використання системних викликів та сервісів операційних систем та їх оточення для розробки нового системного програмного забезпечення;

– забезпечення захищеності програм і даних від несанкціонованих дій;

– верифікація програмного забезпечення;

– розробка інструкцій з використання програм, оформлення необхідної технічної документації.

Аналіз виробничої діяльності дозволив конкретизувати цілі навчання системного програмування бакалаврів програмної інженерії, і на їх основі розробити комплекс практичних завдань з курсу, загальна структура яких представлена в таблиці 1.

*Таблиця 1*

**Структура практичних завдань з дисципліни «Системне програмування»**

| Етапи практичного завдання | Практична діяльність |
|---|---|
| На основі загального формулювання задачі створити технічне завдання на розробку програми | проектування програмного забезпечення |
| Розробити алгоритм програми та відобразити його у вигляді блок-схеми | розробка алгоритмів та структур даних для програмних продуктів |
| Самостійно обрати формати введення початкових даних та виведення результатів | визначення обсягу, структури, схеми введення, опрацювання, збереження і виведення даних, що підлягають обробці засобами обчислювальної техніки |
| Розробити програму згідно отриманому завданню та розробленому алгоритму | конструювання інструментального програмного забезпечення для розробки системного та прикладного програмного забезпечення (компіляторів, текстових процесорів, оболонок операційних систем); конструювання операційних систем та їх оточення; використання системних викликів та сервісів операційних систем та їх оточення для розробки нового |

| Етапи практичного завдання | Практична діяльність |
|---|---|
|  | системного програмного забезпечення; забезпечення захищеності програм і даних від несанкціонованих дій. |
| Протестувати розроблене програмне забезпечення | верифікація програмного забезпечення |
| Підготувати звіт з виконаної практичної роботи | розробка інструкцій з використання програм, оформлення необхідної технічної документації |

Також на стадії проектування були визначені методи навчання, що будуть застосовані під час вивчення окремих тематичних розділів системного програмування. Більшість з цих методів орієнтовані в першу чергу на практичну діяльність майбутніх інженерів-програмістів. Д. Мерріл зазначав, що навчання полегшується, якщо той, хто навчається, займається вирішенням реальних проблем; коли існуючі знання є основою для нових знань; коли той, хто навчається, демонструє і застосовує нові знання; коли нові знання інтегруються у світогляд того, хто навчається [8]. На основі зазначених принципів навчання має складатися з чотирьох основних етапів: активація попереднього досвіду; демонстрація навичок; застосування навичок; інтеграція навичок для вирішення реальних практичних завдань. Ці ідеї найкраще узгоджуються з технологією контекстного навчання, яка дозволяє гармонійно поєднати в собі методи проектів, проблемного навчання, та навчання у співпраці [9]. Саме комбінація цих методів дозволяє найбільш повно розкрити всі компетенції з системного програмування під час навчання.

Для вивчення кожного тематичного розділу з системне програмування з метою активізації пізнавальної діяльності студентів використовуються елементи проблемного навчання, яке передбачає формулювання проблемної ситуації, вирішення якої потребує від того, хто навчається самостійно шукати шляхи її вирішення. Під час виконання практичних завдань, пов'язаних зі сформульованою проблемою, підключаються методи проектів, парного програмування та навчання у співпраці, що дозволяє додатково сформувати необхідні комунікативні професійні навички. У зв'язку з цим, було розроблено наступний план вивчення кожного тематичного розділу:

1. Визначення цілей заняття.
2. Розгляд плану заняття.
3. Постановка проблеми для розв'язання.

4. Генерація ідей, активізація вже отриманих знань та власного досвіду.

5. Узагальнення ідей, що були отримані під час дискусії та розгляд правил і методів, що можна реалізувати на базі цих ідей.

6. Демонстрація практичного застосування розглянутих правил і методів

7. Розв'язання практичних задач в рамках індивідуального або групового проекту.

8. Перевірка засвоєння теоретичного матеріалу.

9. Аналіз, оцінки, висновки.

Деякі етапи такого плану, наприклад, пошукову діяльність пов'язану з генерацією, обговоренням та узагальненням ідей, важко реалізувати в рамках традиційних форм організації навчального процесу – лекції, практичному або лабораторному занятті – у зв'язку з часовими обмеженнями, що мають ці форми занять. Реалізувати ці етапи можна за рахунок більш продуктивного використання часу, що відведено студентам на самостійну роботу, із залученням засобів ІКТ. Найбільш системне використання засобів ІКТ у навчальному процесі вітчизняних ВНЗ досягається за рахунок застосування системи управління комбінованим навчанням, в якості якої може виступати LMS або LCMS, що має достатньо розвинені інструменти комунікації, подання навчальних матеріалів та управління навчальним процесом з урахуванням його комбінованого характеру. Особливості організації роботи студента під час вивчення дисципліни «Системне програмування» з використання системи управління комбінованим навчанням відображено в моделі на рис. 1.

Дана модель передбачає організаційний етап, на якому студент має ознайомитись з особливостями роботи з системою, зареєструватися в системі, отримати доступ до навчальних матеріалів та комунікаційних засобів, що будуть використовуватися під час навчання. На вступному занятті студентам пропонується перевірити навички роботи з системою, після чого можна перейти до вивчення тематичних розділів, яке відбувається за наступним планом: спочатку студентам оголошуються цілі навчання та формулюється проблема, яку належить вирішити в рамках даного розділу.

Оголошення цілей та задач розділу відбувається за допомогою розсилання повідомлень. Після цього починається пошуковий етап, на якому активізується попередній досвід студентів, виконується самостійний пошук методів та засобів вирішення поставленої проблеми, обговорюються та узагальнюються ідеї та обрані методи вирішення задач. На цьому етапі спілкування між студентами та викладачем відбувається за допомогою комунікаційних засобів системи.

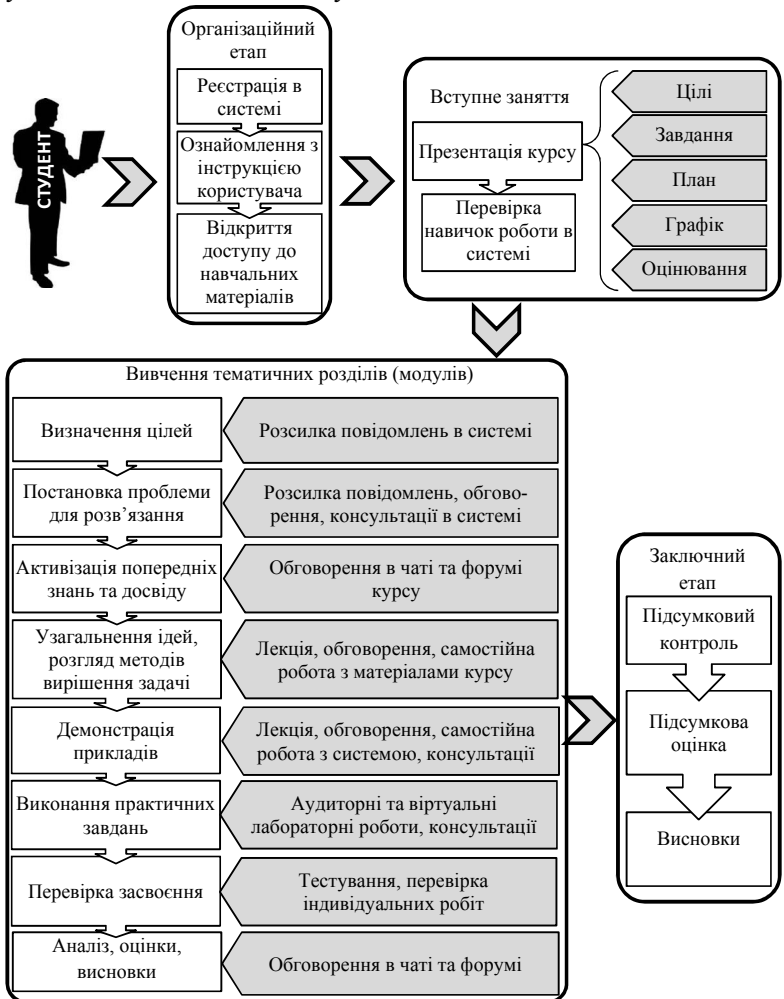

Рис. 2. Модель роботи студента під час вивчення дисципліни «Системне програмування» з використання системи управління комбінованим навчанням

Під час пошуку додаткової інформації студентами використовуються навчальні та інші матеріали, частина яких може бути розміщена в системі. Під час лекційного заняття викладач підводить підсумки пошукового етапу, викладає основні теоретичні відомості, демонструє приклади вирішення поставленої задачі. Після лекційного заняття студентам пропонуються практичні завдання у вигляді індивідуального або колективного проекту, які вони виконують під час лабораторних робіт. Лабораторні роботи можуть виконуватися у спеціально обладнаних аудиторіях, або поза ними з використанням спеціалізованих віртуальних лабораторій у системі. Звіти з виконаної лабораторної роботи подаються на перевірку також з використанням відповідних засобів системи. Після успішного виконання практичних робіт, студенти за допомогою тестування студенти перевіряють рівень засвоєння теоретичних знань.

Завершується вивчення розділу підведенням підсумків, аналізом отриманих результатів та їх обговоренням, що відбувається за допомогою комунікаційних засобів системи.

Результати проектування та розроблена модель роботи студента була використана в Криворізькому національному університеті під час навчання системного програмування студентів спеціальності «Програмне забезпечення автоматизованих систем» з використанням системи управління комбінованим навчанням «Агапа». Але узагальнення результатів проектування дозволяє адаптувати подібну модель до будь якої іншої професійно-орієнтованої дисципліни.

Література